# Terahertz dynamics of electron-vibron coupling in single molecules with tunable electrostatic potential

Shaoqing Du[1, a)], Kenji Yoshida[1], Ya Zhang[1], Ikutaro Hamada[2], and Kazuhiko Hirakawa[1,3,b)]

[1]*Center for Photonics Electronics Convergence, Institute of Industrial Science, University of Tokyo, 4-6-1 Komaba, Meguro-ku, Tokyo 153-8505, Japan*

[2]*Center for Green Research on Energy and Environmental Materials, National Institute for Materials Science, Tsukuba 305-0044, Japan*

[3]*Institute for Nano Quantum Information Electronics, University of Tokyo, 4-6-1 Komaba, Meguro-ku, Tokyo 153-8505, Japan*

Clarifying electronic and vibronic properties at individual molecule level provides key insights to future chemistry, nanoelectronics, and quantum information technologies.  The single electron tunneling spectroscopy[1-4] has been used to study the charging/discharging process in single molecules.  The obtained information was, however, mainly on static electronic properties, and access to their dynamical properties was very indirect.  Here, we report on the terahertz (THz) spectroscopy of single fullerene molecules by using a single molecule transistor (SMT) geometry.  From the time-domain THz autocorrelation measurements, we have obtained THz spectra associated with the THz-induced center-of-mass oscillation of the molecules.  The observed peaks are finely split into two, reflecting the difference in the van der Waals potential profile experienced by the molecule on the metal surface when the number of electrons on the molecule fluctuates by one during the single electron tunneling process.  Such an ultrahigh-sensitivity to the electronic/vibronic structures of a single molecule upon adding/removing a single electron has been achieved by using the THz spectroscopy in the SMT geometry. This novel scheme provides a new opportunity for investigating ultrafast THz dynamics of sub-nm scale systems.

Terahertz (THz) spectroscopy has been developed as a powerful tool for clarifying vibrational dynamics of various kinds of molecules.[5-7] Because of the long-wavelength nature of the THz electromagnetic waves (typically, ~100 μm), however, spectral information was obtained only for an ensemble average of a huge number of molecules. It has been a formidable challenge to by far exceed the diffraction limit and focus the THz radiation on single molecules. Furthermore, the number of mobile charges that can absorb THz radiation in a single molecule is very few, which makes THz absorption extremely small. Here, we report on the THz spectroscopy of single molecules by using the single molecule transistor (SMT) geometry. The SMT is a structure in which a single molecule is captured in a sub-nm gap created between the source and drain metal electrodes fabricated on a field plate. Using the source and drain electrodes separated by a sub-nm gap as a THz antenna, we focus THz radiation onto a single fullerene molecule trapped in the nanogap electrodes. Furthermore, we can detect a very small absorption by measuring the THz-induced photocurrent by the same electrodes.

We fabricated $C_{60}$ SMTs by using the electrical break junction (EBJ) method.[8,9] Figure 1a shows a scanning electron microscope (SEM) image of the sample structure used in this work. A 10-nm-thick NiCr layer, which serves as a semi-transparent backgate electrode, was deposited on a high-resistivity Si substrate, and a 30-nm-thick $Al_2O_3$ gate-insulation film was grown by atomic layer deposition. We formed thin gold nanojunctions, which were used for forming the source and drain electrodes, on the surface of the wafer by the shadow evaporation technique. To achieve a good coupling efficiency between the THz radiation and the single molecule, we employed a bowtie-antenna shape for the source and drain electrodes, as shown in Fig. 1a. Since the fabrication yield of SMTs is typically only a few %, we fabricated 9 junctions on a chip in an area of $200 \times 200$ μm$^2$. The chip was then glued with varnish on a hemispherical silicon lens to tightly focus the THz radiation on the bowtie antenna. A dilute toluene solution of $C_{60}$ molecules was deposited on the surface of the gold nanojunctions and dried off with nitrogen gas. The sample was then mounted in a vacuum space of a $^4$He cryostat. We applied the feedback-controlled EBJ process to the metal nanojunctions and fabricated source/drain electrodes separated by a sub-nm gap, as shown in Fig. 1b.[8-10] Details of the feedback control of the EBJ process have been described elsewhere.[10] All the measurements were performed at ~4.6 K in a vacuum.

By using the nanogap electrodes[11-13], THz radiation can be tightly focused onto a single molecule. Furthermore, the local THz field in the nanogap is enhanced by a factor of ~$10^5$ by the plasmonic effect of the metal electrodes.[13-15] Here, we face a problem in performing spectroscopy on a single molecule by using broadband THz bursts excited by femtosecond laser pulses; the SMTs are very slow devices due to their high tunnel resistances and it is impossible to read ultrafast current changes. To overcome this problem, we adopted the time-domain THz autocorrelation measurements,[16-19] as shown in Fig. 1c. Using a beam splitter, we split the laser beam into two parts and created double laser pulses. A surface of an InAs wafer was consecutively pumped by the femtosecond laser pulses and the time-correlated THz double pulses were generated. By recording the photocurrent induced by the THz double pulses as a function of the time interval between the THz pulses, $\tau$, we obtained interferograms of the photocurrent induced in the single molecule by THz radiation (quasi-autocorrelation measurement; see METHODS for more details). We used a mode-locked Ti:sapphire laser with a central wavelength at 810 nm, pulse duration ~10 fs, and repetition frequency of 76 MHz. The output power of the laser was typically 500 mW. The generated THz waveform was measured by electrooptic (EO) sampling in a 700 μm-thick (110) ZnTe crystal. As shown in Fig. 1d, the THz pulses had a monocycle waveform and their peak THz field was approximately 0.4 V/cm. Since the bandwidth of the 700 μm-thick EO sensor is limited up to 3 THz, we characterized the emitted THz radiation also by using a wideband Si bolometer in the quasi-autocorrelation geometry. As shown in Fig. 1e, the power spectrum of the emitted THz radiation extends up to 12 THz.

We performed transport measurements to examine whether we captured a single molecule in the nanogap (see Supplementary Information for the sample statistics). Figure 2a shows a color-coded differential conductance ($\partial I_{SD}/\partial V_{DS}$) map of a $C_{60}$ SMT plotted as a function of the source-drain voltage, $V_{DS}$, and the gate voltage, $V_G$ (Coulomb stability diagram). A crossing pattern in the diagram indicates that this device operates as a single electron transistor. Furthermore, the excitation of an internal vibrational mode at ~33 meV,[20] which is intrinsic to the $C_{60}$ molecule, was observed in the Coulomb stability diagram, confirming that a single $C_{60}$ molecule was indeed trapped in the nanogap of the electrodes and served as a Coulomb island.

After identifying the molecule from transport measurements, we illuminated the sample with THz pulses by setting $\tau \gg 10$ ps and searched for the time-averaged THz induced photocurrent. By sweeping $V_G$ while applying a small source-drain voltage ($V_{DS} = 0.1$ mV), we measured the THz-induced photocurrent as a function of $V_G$. The black curve in Fig. 2b is the sample conductance in the dark condition, while the red trace is the measured THz-induced photocurrent. The peak of the black curve corresponds to the charge degeneracy point of the $C_{60}$ SMT. A very small, but finite photocurrent of the order of 100 fA was observed near the charge degeneracy point. Note that Cocker et al. recently performed single molecule THz spectroscopy by using a scanning tunneling microscope (STM) tip to focus the THz radiation onto a single pentacene molecule.[21] Although they can perform site-selective THz spectroscopy by imaging the molecular orbitals by STM,[21] the great advantage of our single molecule transistor geometry is that we can precisely control the electrostatic potential and the number of electrons on the molecule by the gate electric fields. Another important difference is that, since the incident THz field is weak (< 0.4 V/cm) in our experiment, the effect of instantaneous bias change induced across the $C_{60}$ molecule is negligibly small (in the order of 1 mV). Therefore, the present experiment was performed in a regime very different from that for the work by Cocker et al.[21] and provides information complimentary to theirs.

Figure 2c shows an interferogram of the THz-induced photocurrent measured for a $C_{60}$ SMT when the gate voltage was set at the peak position of the photocurrent ($V_G = 0.024$ V). The red curve is the interferogram obtained by averaging 10 scans. A clear center peak and interference feature can be seen. By calculating the Fourier spectrum of the interferogram, we obtained a THz spectrum for a $C_{60}$ SMT. Two peaks are observed at around 2 meV and 4 meV. We analyzed the inerferogram shown in Fig. 2(c) and decomposed the waveform into the 2 meV (500 GHz)- (blue curve) and 4 meV (1 THz)- (green curve) oscillation components, as shown in Fig. 2d. The oscillation of the 4 meV-component is clearly visible in the photocurrent waveform, particularly, at around 1 ps. It should be noted that a THz vibrational spectrum is obtained even for a single molecule when the SMT geometry is used. Similar sharp peaks were observed also for an endohedral metallofullerene Ce@$C_{82}$ SMT, as shown in Fig. 2e.

The low-energy excitations observed for a $C_{60}$ SMT originate from the vibron-assisted tunneling promoted by the THz-induced center-of-mass oscillation of the $C_{60}$ molecule. Park et al. observed vibron-assisted inelastic tunneling in the tunneling spectroscopy measurements.[1] In the transport experiments, the molecular vibration is excited by injecting tunnel electrons and the excited state lines are visible only inside the single electron tunneling region of the Coulomb stability diagrams. In the present THz spectroscopy, molecular vibrations are excited not only by the tunneling electrons but also by the impulsive THz fields generated by the femtosecond laser pulses. Because of the broadband THz excitation, the center-of-mass oscillation of the $C_{60}$ molecule is excited and creates new tunneling paths for electrons; *i. e*., the vibron-assisted tunneling processes as schematically shown in Figs. 3a and 3b. When the lowest unoccupied molecular orbital (LUMO) level is above the Fermi level of the electrodes, an electron in the electrodes cannot enter the molecule in the dark condition. However, when the center-of-mass oscillation of the $C_{60}$ molecules is excited by the THz radiation, an electron in the electrode can absorb a vibron and tunnel into the molecule (vibron-assisted tunneling). In this case, the electron number on the molecule increases from *N* to *N*+1; *i. e*., the molecule changes its state from $C_{60}$ to $C_{60}^-$. Although *N* cannot be determined only from the present experiment, *N* is most likely to be zero. Figure 3b shows a vibron-assisted tunneling-out process when the highest occupied molecular orbital (HOMO) level is below the Fermi level of the electrode. In this case, the electron on the HOMO level cannot leave the molecule in the dark condition, but, by THz illumination, it can escape the molecule via the vibron-assisted tunneling. Furthermore, we can roughly estimate the quality (Q)-factor of the $C_{60}$ molecular vibration to be approximately 3-5, although the linewidths of the sharp peaks are affected by the resolution of the experimental setup. This means that an electron that hops on the $C_{60}$ molecule resides on the molecule at least for about 3-5 cycles of vibration and leaves the molecule. We have estimated the electron tunneling time, $\tau_T$, through the $C_{60}$ molecule to be 8 ps from the sample conductance in the dark condition (~7 μS), using the following relationship;[22]

$$G_{max} = \frac{e^2}{4k_B T} \frac{G^l G^r}{G^l + G^r},\qquad(1)$$

and assuming $\Gamma^l \approx \Gamma^r \approx 1/\tau_T$. Here, $e$ is the elementary charge, $k_B$ the Boltzmann constant, $T$ the temperature, and $G_{max}$ the conductance of the Coulomb peak. $\Gamma^l$ ($\Gamma^r$) is the tunnel coupling between the molecule and the left (right) electrode. The obtained $\tau_T$ is consistent with the electron dwell time on the molecule determined from the Q-factor.

A very interesting observation is that the vibron peaks around 2 meV and 4 meV are finely split into two. The magnitudes of the peak splittings at around 2 meV and 4 meV are 0.8 meV and 0.6 meV, respectively. We measured five $C_{60}$ SMTs in total and four of them showed the same splitting feature. In addition, a similar peak splitting was observed in the THz spectrum of a Ce@$C_{82}$ SMT. The excited-state lines due to molecular vibrations in the Coulomb stability diagrams originate from the Franck-Condon effect.[23-25] As schematically shown in Fig. 3d, the overlap of the vibrational wave functions gives the tunneling probability of an electron for the transition between the $N$- and $N+1$-states (the Franck-Condon principle).[23-25] When the charge-state changes from the $N$- to $N+1$-electron state, the equilibrium position of the molecule may shift by $\delta$. This shift induces not only the diagonal transitions but also the off-diagonal transitions between the vibrational states of the $N$- and $N+1$-charge states, giving rise to multiple excited-state lines in the Coulomb stability diagrams.[25] (see Supplementary Information) In the previous discussions[1,23-25], the vibrational frequencies of the molecule for the *N*- and *N*+1-charge states were assumed to be the same. However, this may not be the case in actual systems; *i.e.*, the van der Waals potential felt by a $C_{60}$ molecule on the gold surface may depend on the charge state of the molecule.

To gain insight into the vibrational and electronic states of the $C_{60}$ molecule in the SMT geometry, we performed van der Waals inclusive[26] density functional theory (DFT) calculations of neutral and negatively charged $C_{60}$ on a Au(111) surface, which correspond to the *N*- and *N*+1-charged states, respectively. The method of the calculation is described in the supplementary information. In Fig. 4, we plot the interaction energy curves for neutral and negatively charged $C_{60}$ on a Au(111) surface. The equilibrium $C_{60}$-surface distance is 0.243 nm and the vibrational energy, $\hbar\omega$, is 4.1 meV. The calculated vibrational energy is larger than the value obtained by the THz spectroscopy (~ 2 meV), presumably because the counter electrode is lacking in the present calculation. We have found that, when $C_{60}$ is negatively charged, the system gets destabilized, because the antibonding state between the LUMO of $C_{60}$ and the substrate state is partially occupied.[27] Accordingly, the equilibrium distance becomes longer by 0.011 nm and the vibrational energy is lowered by 0.7 meV. This result is opposite to a simple expectation when only the image-charge force in the metal electrode is considered. Our calculations suggest that the vibrational energy depends on the charge state and, thus, the splitting of the vibron-assisted tunneling peak is expected. Indeed, the magnitude of the observed splitting (0.6~0.8 meV) is in good agreement with the calculated change in the vibrational energy (0.7 meV).

Here, we would like to add a comment on why we observe the vibron peaks for the *N*- and *N*+1-charge state at the same time in the THz spectrum. As seen in Fig. 2b, the peak of the THz-induced photocurrent is located on the negative-side tail of the Coulomb peak ($V_G$ = 0.024 V). At this gate voltage, the LUMO is estimated to be 4 meV above the Fermi level, using the lever-arm factor, $\alpha$ = 0.03, in this sample. On the other hand, as shown in Fig. 2a, the tunnel peak of this SMT is very broad (~10 meV), much broader than the energy difference between the Fermi level and the LUMO.

Therefore, an electron can hop on to or hop out of the molecule at this bias condition and the number of electrons on the molecule can fluctuate (either $N$ or $N+1$), as schematically illustrated in Fig. 3c.

In summary, we have demonstrated that THz spectroscopy can detect an ultrafast oscillatory motion of a single $C_{60}$ molecule. Low-energy vibrational modes are observed at around 2 meV and 4 meV and are attributed to the THz-induced center-of-mass nanomechanical oscillation of the $C_{60}$ molecule. Furthermore, we have found that the observed THz peaks are finely split into two, reflecting the difference in the van der Waals potential profile experienced by the $C_{60}$ molecule on the metal surface when the number of electron on the molecule fluctuates between $N$ and $N+1$ during the single electron tunneling process. Such an ultrahigh-sensitivity to the electronic/vibronic structures of a single molecule upon charging/discharging a single electron has been achieved by using metal source-drain electrodes separated by a sub-nm gap as a THz antenna and detecting the THz-induced photocurrent.

**Acknowledgements**

We thank Y. Arakawa for his continuous encouragement and S. Ishida for his technical support in the fabrication process. This work is supported by MEXT KAKENHI on Innovative Areas "Science of hybrid quantum systems" (15H05868), KAKENHI from JSPS (16H06709 and 17H01038), and Project for Developing Innovation Systems of MEXT.

**Contributions**

S.Q.D. fabricated the single molecule transistor samples and carried out the terahertz measurements. K.H. conceived and supervised the project. K.Y. supported the transport measurements and Y.Z. provided assistance in the THz spectroscopy. I.H. carried out the DFT calculations. S.Q.D., I.H. and K.H. wrote the manuscript with contributions from all authors. All authors contributed to discussions.

**References**

a) Electronic mail: sqdu@iis.u-tokyo.ac.jp

b) Electronic mail: hirakawa@iis.u-tokyo.ac.jp

**Figure Captions**

**Figure 1   Setup for single molecule terahertz spectroscopy.**   **a**, SEM image of a SMT sample with a bowtie antenna structure.   S, D, and G denote the source, drain, and gate electrodes of the SMT, respectively.   **b**, SEM images of the nanojunction region before and after the electrical break-junction process.   A single molecule is captured in the created nanogap.   **c**, Schematic illustration of the experimental setup.   Femtosecond laser pulses (center wavelength: 810 nm, pulse duration: ~10 fs, repetition rate: 76 MHz) are split into two parts and focused on a THz emitter (InAs wafer).   The generated THz bursts are collected and focused onto the single molecule transistor (SMT) mounted in a $^4$He cryostat.   **d,** THz waveform used in the measurements.   THz pulses were generated by pumping the surface of an InAs wafer by 10 fs laser pulses from a mode-lock Ti:sapphire laser.   The THz waveform was measured by using the electro-optic (EO) sampling method.   The EO sensor used in the measurement was a 700 μm-thick (110) cut ZnTe.   The THz pulse has a monocycle waveform and the maximum THz field ~ 0.4 V/cm.   **e,** Spectrum of the THz pulses measured by using a wideband Si bolometer in the quasi-autocorrelation geometry (see METHODS).   The THz spectrum extends up to 12 THz.   Dips observed in the spectrum in a range from 1-5 THz are due to water vapor absorption.

**Figure 2   Teraherz autocorrelation measurements of a single $C_{60}$ molecule.**   **a**, Coulomb stability diagram of a $C_{60}$ SMT.   White dashed lines are eyeguides for the boundaries of the Coulomb diamonds.   The excited state line at 33 meV is an internal vibrational mode of the $C_{60}$ molecule.   **b**, The single electron tunneling current $I_{DS}$ (black curve) and the THz-induced photocurrent $I_P$ (red curve) as a function of $V_G$ measured at $V_{DS} = 0.1$ mV.   **c**, Quasi-autocorrelation trace (interferogram) of the THz induced photocurrent measured at the peak of the photocurrent ($V_G = 0.024$ V, $V_{DS} = 0.1$ mV).   The quasi-autocorrelation trace was obtained by averaging data for eight scans.   **d,** The interferogram for $\tau > 0$, replotted from **c**.   Dots are the measured data points.   The blue and green curves are fitting curves for the 500 GHz- and 1 THz- oscillation components, respectively.   The peak splitting discussed later was not included in the fitting.   **e,** Fourier spectrum of the interferogram shown in **c**.   The spectrum exhibits sharp peaks at ~2 meV and ~4 meV.   Additional

data for other samples are shown in Supplementary Information.  **f**, THz spectrum for a single-Ce@$C_{82}$ SMT.  Two sharply split peaks are observed at around 2 meV.

**Figure 3  Vibron-assisted tunneling in a $C_{60}$ single molecule transistor.**  **a**, Tunnel-in process: when the lowest unoccupied molecular orbital (LUMO) level is above the Fermi level of the electrodes, an electron in the electrode can absorb a vibron and jump into the molecule when molecular vibration is excited by THz.  **b,** Tunnel-out process: when the highest occupied molecular orbital (HOMO) level in the molecules is below the Fermi level of the electrode, the electron can absorb a vibron and jump out of the molecule when molecular vibration is excited by THz.  **c**, The energy band diagram at the peak of the photocurrent ($V_G$ = 0.024 V, $V_{DS}$ = 0.1 mV) in Fig. 2b.  The Lorentzian curve schematically illustrates the electronic level in the molecule.  The broadening of the LUMO level, $\Gamma$, is determined to be ~10 meV from the Coulomb stability diagram shown in **a**.  The center of the LUMO level is located 4 meV above the Fermi levels in the electrodes ($\delta E$).  Since $\Gamma$ is larger than $\delta E$, not only the vibron-assisted tunnel-in process but also tunnel-out process are possible.  **d**, Schematic illustration of the vibronic energy diagram of the $C_{60}$ molecule in the nanogap.  The $C_{60}$ molecule experiences the van der Waals potential on the surface of the gold electrode and performs the center-of-mass oscillation.  The parabolas and the wavefunctions in the diagram schematically illustrate the vibronic energy states for the $N$- and $N$+1-electron states.  When an electron is added to the $C_{60}$ molecule, the equilibrium position of the molecule shifts by $\delta$.  The vibronic energies for the $N$- and $N$+1-electron states are expressed as $\Delta E(N)$ and $\Delta E(N+1)$, respectively.  Due to the resonant absorption of THz photons, electron tunneling via the GS($N$)↔ES($N$+1) and ES($N$)↔GS($N$+1) transitions become possible and generate THz-induced photocurrent even in the Coulomb gap, where GS and ES denote the ground state and the excited state, respectively.

**Figure 4  Calculated interaction energy for a $C_{60}$ molecule on a gold surface.**  Open diamond and filled diamond display the interaction energies calculated as a function of the molecule-surface distance for $C_{60}$ and $C_{60}^-$ on a Au(111) surface, respectively.  The molecule-surface distance is defined as the difference between the average $z$-coordinates of the bottom carbon atoms of $C_{60}$ and the surface gold atoms, as denoted by an arrow in the inset. For the $C_{60}$ molecule on the gold surface (the $N$-electron state), the characteristic vibronic energy, $\hbar\omega$, is 4.1 meV. When an electron is added to the $C_{60}$ molecule (the $N$+1-electron state), the equilibrium distance becomes larger by 0.011 nm. At the same time, the characteristic vibronic energy becomes smaller by 0.7 meV.

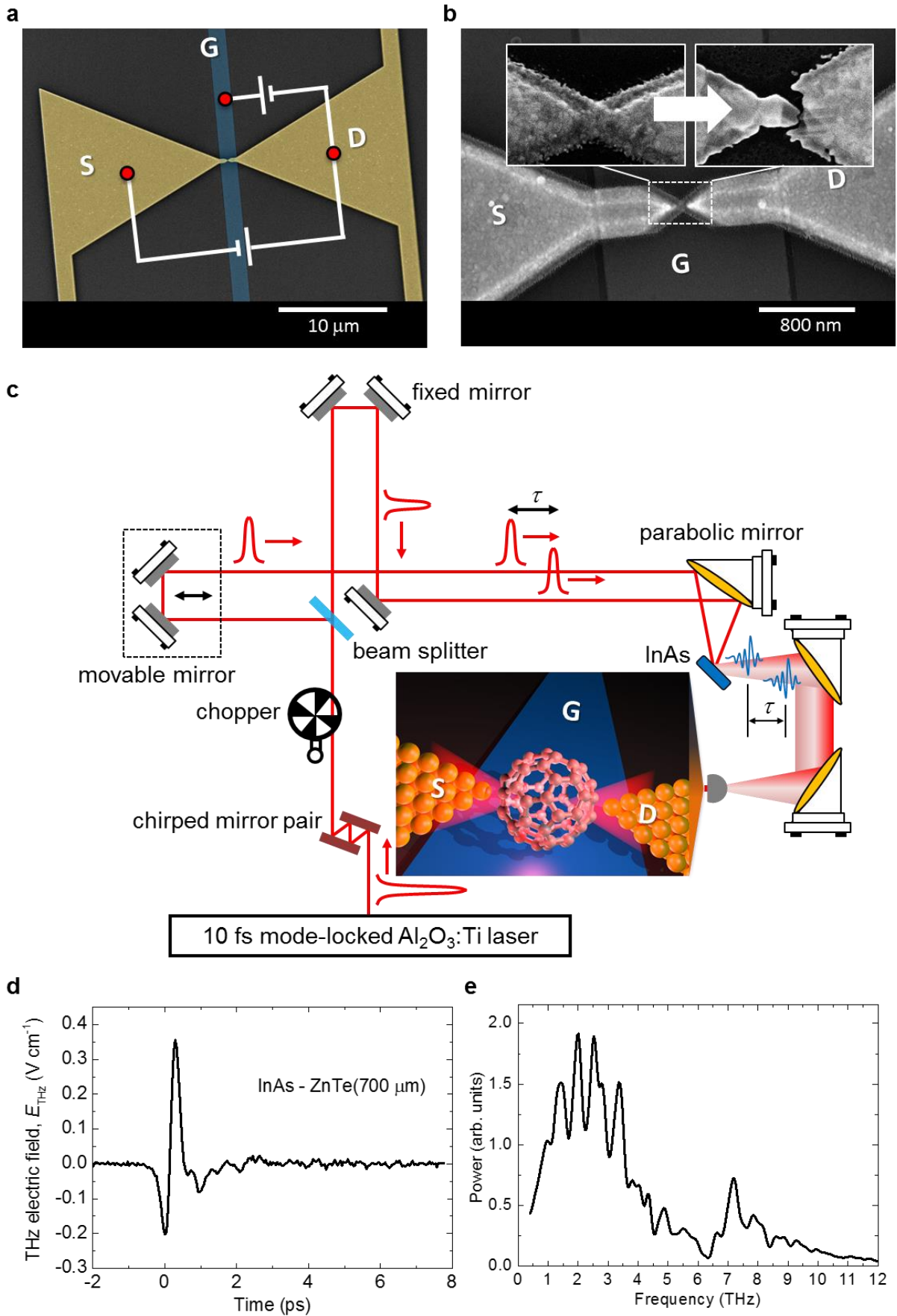

Fig. 1    S. Q. Du, et al.

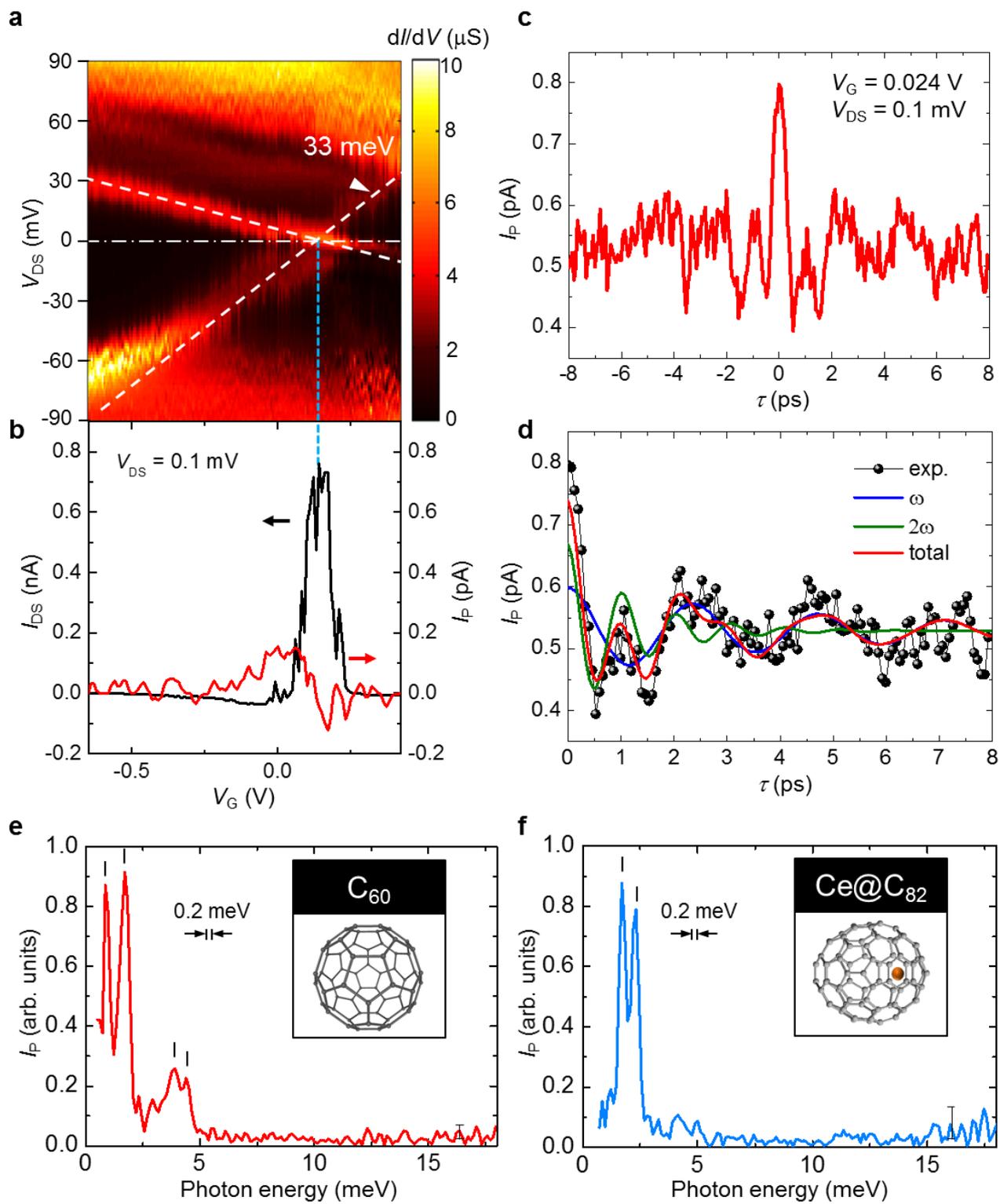

Fig. 2   S. Q. Du, et al.

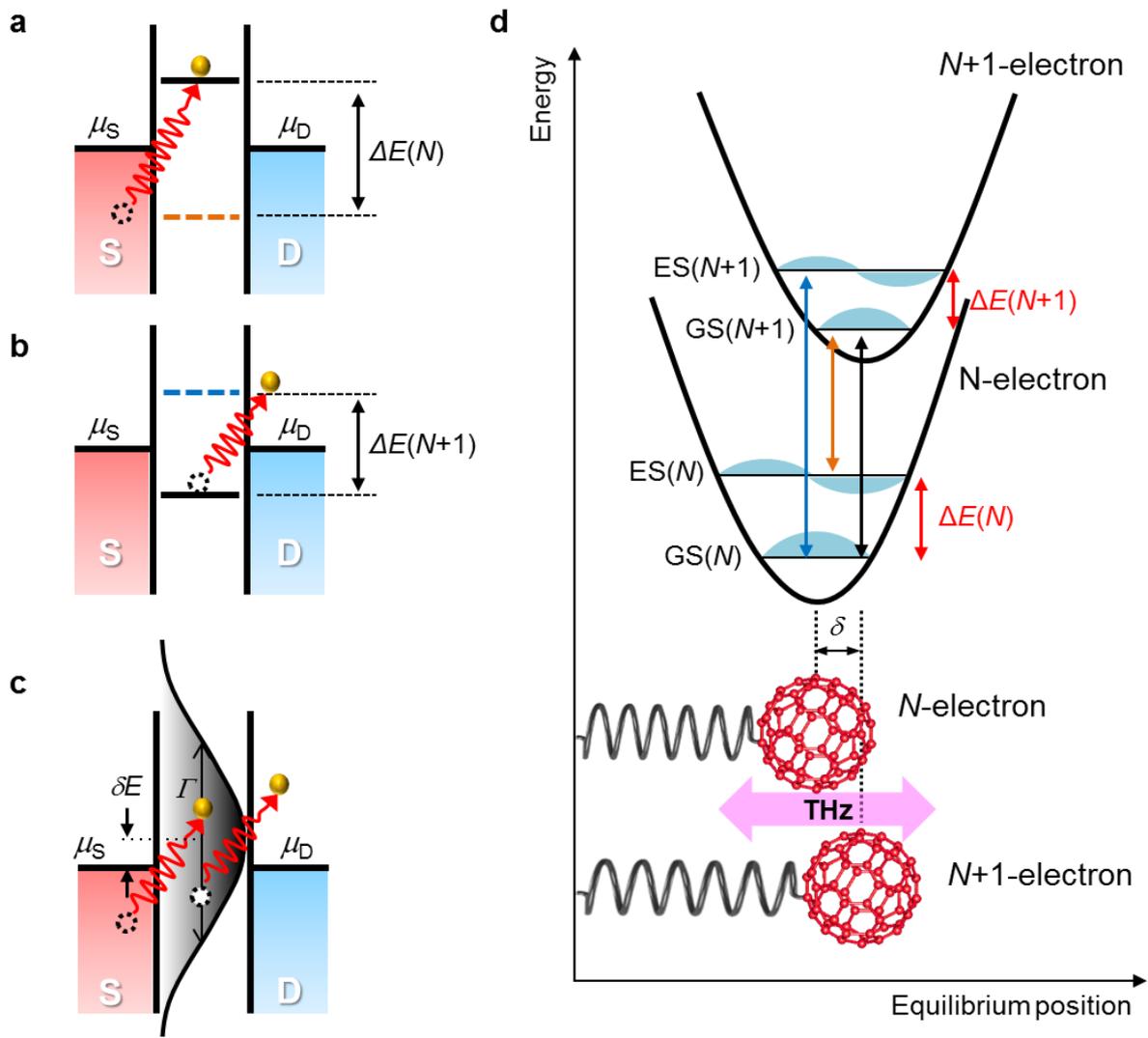

Fig. 3  S. Q. Du, et al.

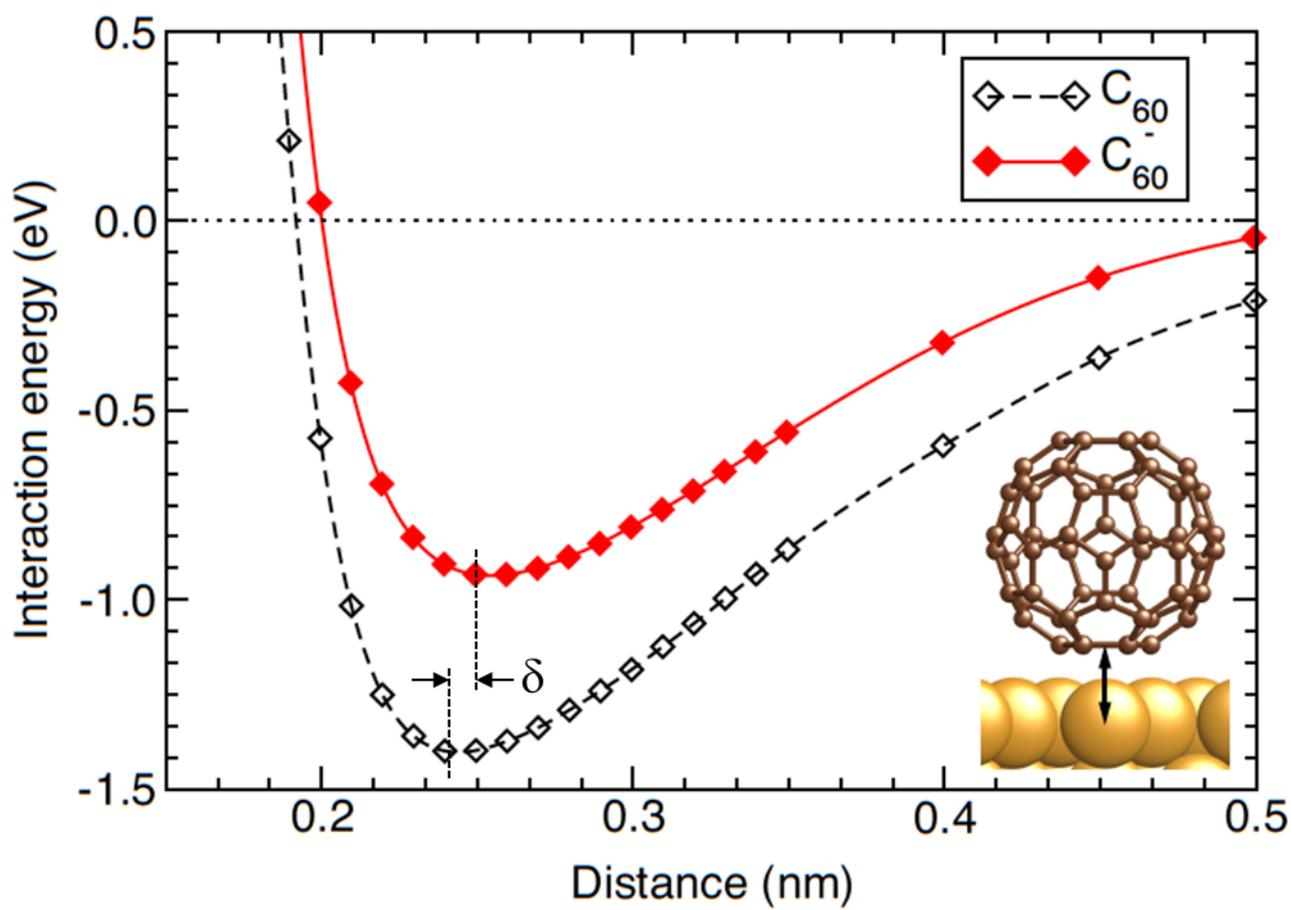

Fig. 4   S. Q. Du, et al.

**METHODS**

**Quasi-autocorrelation measurements.** The standard THz auto-correlation measurements are performed by splitting the THz radiation into two parts and recombine them by using, for example, a Michelson interferometer and measuring the power of the THz radiation. In this work, we have used a slightly different method (quasi-autocorrelation measurement[16,17]). By using an optical beam splitter, we first split the excitation laser beam into two parts and created double laser pulses. A surface of an InAs wafer was consecutively pumped by the femtosecond laser pulses and the time-correlated THz double pulses were generated. We used a mode-locked Ti:sapphire laser with a central wavelength at 810 nm, pulse duration ~10 fs, and repetition frequency of 76 MHz for excitation. The typical output power of the femtosecond laser was 500 mW. By recombining the double THz pulses and measuring the total THz power as a function of the time interval between the two THz pulses, $\tau$, we obtained interferograms of the THz radiation shown in Extended Data Figure 1.

If the surface of an InAs wafer is consecutively excited with double fs laser pulses exactly at the same position, unwanted artifacts such as interference fringes, bleaching, etc. may be induced. In order to avoid such artifacts, we carefully shifted the positions of the laser excitation in such a way that 1) we do not observe interference fringes in the interferograms, 2) the ratio between the center peak and the background is as close to 2 as possible, and 3) the obtained interferogram is as symmetric as possible, as shown in Fig. 1. By adjusting the optics in such a way, we can generate time-correlated THz double pulses with almost identical waveforms.

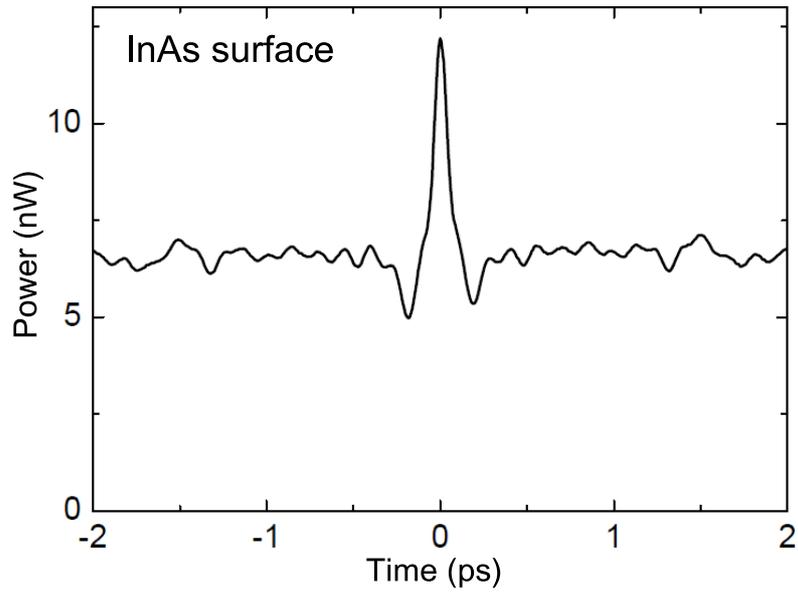

**Extended Data Figure 1** | Quasi-autocorrelation trace (interferogram) of the THz radiation generated from an InAs surface measured by using a broadband Si bolometer.

**Source-drain electrodes with a terahertz bowtie-antenna geometry.** We have fabricated the source and drain electrodes of the single molecule transistors (SMTs) in a shape of a bowtie antenna in order to enhance the coupling efficiency between the THz radiation and the single molecules. The total antenna length was 33 μm and the angle of the bowtie was set to be 57°. Extended Data Figure 2a shows the antenna design pattern and a color-coded THz field strength calculated at 1.6 THz. As seen in the figure, the radiation at 1.6 THz is enhanced in the nanogap region. Extended Data Figure 2b shows the resonance spectrum of the antenna calculated by using the finite element method. The antenna has a broad resonance feature around 2 THz and covers a frequency range from 100 GHz up to 4 THz.

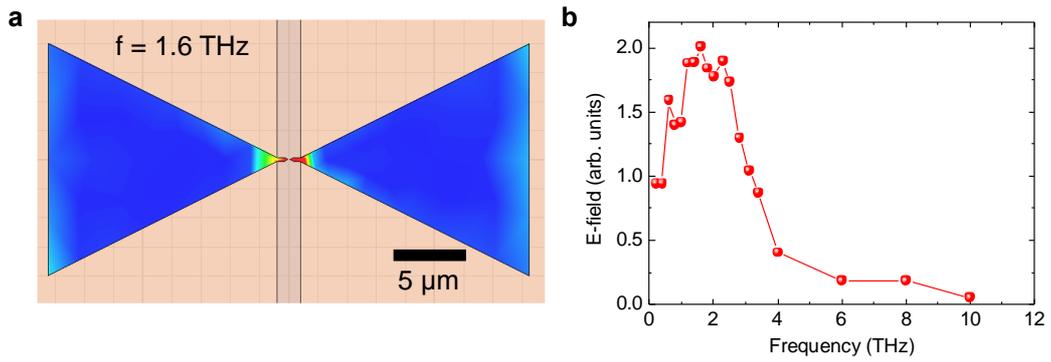

**Extended Data Figure 2** | **a.** Top view of the source/drain electrodes used in the present work. The electric field distribution at 1.6 THz is plotted in a color code. As seen in the figure, the terahertz field is localized in the gap region. **b.** Spectrum of the THz field strength in the nanogap region. The antenna has a broad resonance around 2 THz.